\documentclass[aip,aps,preprint,showpacs]{revtex4}
\usepackage{amssymb}

\usepackage{graphicx}

\usepackage[T1]{fontenc}

\usepackage[american]{babel}

\begin{document}

\title{Tunable striped-patterns by lattice anisotropy and magnetic impurities in
d-wave superconductors}

\author{Xian-Jun~Zuo$^{1}$}\thanks{xjzuo@yahoo.com.cn}
\author{Yuan~Zhou$^{1}$}
\author{Chang-De~Gong$^{2,1}$}
\affiliation{$^1$National Laboratory of Solid State Microstructures and
Department of Physics, Nanjing University, Nanjing 210093,
People's Republic of China\\
$^2$Center for Statistical and Theoretical Condensed Matter Physics,
Zhejiang Normal University, Jinhua 321004, People's Republic of
China}

\date{\today}

\begin{abstract}
\quad The pattern transition induced by lattice anisotropy (LA) and magnetic
impurities is computationally observed in near-optimally doped
d-wave superconductors (DSCs). For the single impurity
case, a transition from the checkerboard to stripe pattern can be
induced even with a very weak LA. Moreover,
the modulation period of eight lattice constants (8$a$) in the spin
order coincides with neutron scattering data. For the two-impurity
case, an orientation transition from the longitudinal
impurity-pinned stripe into the transverse pattern is observed when
the LA ratio reaches some critical value. At the
critical point, it is found that the structures around magnetic
impurities could restore checkerboard patterns. These results
indicate that the formation of stripes in DSCs might induced by
various effects, and could be tunable experimentally.

\end{abstract}


\maketitle

\bigskip



The inhomogeneous phases in unconventional superconductors have
attracted much attention recently. Various experiments
reported the presence of stripe or checkerboard modulations in
copper oxide-based
compounds~\cite{Tranq1,mook2,fujita3,lake4,hoff5,howa6,vers7,hana8,Mc9,Kohs10,khay11,jfding12}.
Neutron scattering (NS) measurements on cuprates such as
La$_{2-x-y}$Nd$_{y}$Sr$_{x}$CuO$_{4}$ (LNSCO),
La$_{2-x}$Sr$_{x}$CuO$_{4}$ (LSCO), La$_{2-x}$Ba$_{x}$CuO (LBCO),
and Y-Ba-Cu-O (YBCO) family indicated the existence of
incommensurate magnetic peaks, which has led to discussions of the
existence of a stripe phase~\cite{Tranq1,mook2,fujita3}.
Several scanning tunneling microscopy (STM) experiments observed
checkerboard-like charge-density wave (CDW) modulations with a
period of roughly four lattice constants (4$a$). Experimentally,
both dispersive~\cite{hoff5} and
non-dispersive~\cite{howa6,vers7,hana8} modulated patterns have been
observed in cuprates. The corresponding patterns were proposed to be
understood in terms of quasiparticle interference
(QPI)~\cite{zhujx13,wangqh14}, or in terms of static or fluctuating
stripes~\cite{kivelson15}, respectively. Currently, the issue of the
nature of these modulated patterns is still under debate.

It is known that lattice anisotropy (LA) are ubiquitous in various
cuprates. For instance, strong a-b axis asymmetry of both the normal
and superconducting state electronic properties have been observed
in the YBCO family~\cite{basov16,Lu17,kondo18}. Up to now, the
relationship among lattice distortions, incommensurability and
stripes has been intensively studied in
cuprates~\cite{normand19,yang20,emery21,becca22,arai23,yexs24,eremin25,hinkov26}.
Based on the anisotropic Hubbard ($t_{x}\neq t_{y}$) and
$t_{x}-t_{y}-J_{x}-J_{y}$ models, Normand and Kampf et al.
considered LA as a possible origin of the stripe
formation in the cuprate superconductors~\cite{normand19}. Moreover,
Becca et al. have shown that stripes and spin incommensurabilities
are favored by LA ~\cite{becca22}. In this work,
we explore the effect of LA on the patterns around
magnetic impurities in d-wave superconductors (DSCs). Including the
competition and coexistence among the DSC, spin, and charge orders,
we study the system by self-consistently solving the Bogoliubov - de
Gennes (BdG) equations based on an anisotropic Hubbard-type model.


We start from the two dimensional $t_{x}-t_{y}-t^{\prime}-U-V$
model, which consists of two parts, $H=H_{0}+H_{imp}$. The
Hamiltonian $H_{0}$ and $H_{imp }$ describe the superconductor and
magnetic impurities, respectively, which is given by
\begin{eqnarray}
H_{0}&=&-\sum_{ij\sigma}t_{ij}(c_{i\sigma}^{\dagger}c_{j
\sigma}+H.c.)  \nonumber \\
& +&\sum_{ij}(\Delta_{ij}c_{i\uparrow}^{\dagger}c_{j\downarrow}^{
\dagger}+H.c.)+\sum_{i\sigma}(Un_{i,\bar{\sigma}}-\mu)c_{i\sigma}^{\dagger}c_{i\sigma},
\nonumber
\\
H_{imp}&=
&\sum_{i}h_{eff}(i)(c_{i\uparrow}^{\dagger}c_{i\uparrow}-c_{i\downarrow}^{
\dagger}c_{i\downarrow}),
\end{eqnarray}
Here $c_{i\sigma}$ annihilates an electron of spin $\sigma$ at the
$i$th site. The hopping integral $t_{ij}$ takes $t_{x}$ or $t_{y}$
between nearest neighbor (NN) pairs along x or y direction, and
$t^{\prime}$ between next-nearest neighbor (NNN) pairs. $U$ is the
on-site Coulomb repulsion interaction. $\mu$ is the chemical
potential, which is determined self-consistently in the calculation.
The local effective field $h_{eff}(i)$ is introduced to model the
exchange coupling between conducting electrons and the impurity
spin, where we have treated the impurity spin as a Ising-like one.
Some similar model had been employed to study the effects of
magnetic impurities on cuprate superconductors~\cite{bala27,zuo28},
which can qualitatively explain the observed impurity states
well~\cite{huds29}. Therefore, we employed the above model in this
work. Experimentally, the ratio of the lattice constants along x and
y directions is $b/a\sim1.01$ (Ref.~\cite{jorgen30}), thus the
corresponding effective hopping integrals $t_{x}$ and $t_{y}$ are
also anisotropic. According to the local-density approximation band
calculation~\cite{andersen31}, the ratio of hopping integrals is
estimated as $t_{x}/t_{y}\sim(b/a)^{4}\sim1.04$. In this work, we
consider effects of LA on the patterns around
magnetic impurities by tuning the ratio $\eta=t_{y}/t_{x}$, which is
near the above estimated value, as shown below. The self-consistent
mean-field parameters are given by $n_{i}=\sum_{\sigma}
<c_{i\sigma}^{\dagger}c_{i\sigma}>$, the magnetization $m_{i}=(1/2)(<c_{i%
\uparrow}^{\dagger}c_{i\uparrow}>-<c_{i\downarrow}^{\dagger}c_{i\downarrow}>)
$, and the DSC order parameter
$\Delta_{ij}=(V/2)<c_{i\uparrow}c_{j\downarrow}-c_{i\downarrow}c_{j\uparrow}>$
with V the phenomenological pairing interaction.

The Hamiltonian $H$ can be diagonalized by solving the following BdG
equations,
\begin{eqnarray}
\left(
\begin{array}{lr}
H_{ij,\uparrow} & \Delta_{ij} \\
\Delta_{ij}^{\ast} & -H_{ij,\downarrow}^{\ast}
\end{array}
\right)\Psi_{j}=E\Psi_{i},
\end{eqnarray}
where the quasiparticle wave function $\Psi_{i}=(u_{i\uparrow},v_{i
\downarrow})^{T}$. The spin-dependent single-particle Hamiltonian
reads
$H_{ij\sigma}=-t_{x}\delta_{i+\hat{x},j}-t_{y}\delta_{i+\hat{y},j}-t^{\prime}\delta_{i+
\tau^{\prime},j}+[\sum_{i_{m}}\sigma
h_{eff}(i)\delta_{i,i_{m}}+Un_{i,\bar{\sigma}}-\mu]\delta_{ij}$.
Here the subscripts $\hat{x}$ and $\hat{y}$ denote the unit vector
directing along x- or y-direction NN bonds. $ \tau^{\prime}$ denotes
the unit vector directing along four NNN bonds, and $i_{m}$ is the
position of the impurity site. The self-consistent parameters are
given by $n_{i\uparrow}=\sum_{n}|u^{n}_{i \uparrow}|^2f(E_{n})$,
$n_{i\downarrow}=\sum_{n}|v^{n}_{i \downarrow}|^2[1-f(E_{n})]$, and
$\Delta_{ij}=\frac{V}{4}\sum_{n}[{
u^{n}_{i\uparrow}v_{j\downarrow}^{n\ast}+v_{i\downarrow}^{n\ast}u^{n}_{j
\uparrow}}]tanh(\frac{\beta E_{n}}{2})$, where $f(E)=1/(1+e^{\beta
E})$ is the Fermi-Dirac distribution function. Hereafter, the length
is measured in units of the lattice constant $a$, and the energy in
units of $t=t_{x}$. The pairing interaction is chosen as $V=1.0$ to
guarantee that the superconducting order $\Delta_{0}\simeq0.08t$,
comparable with the observed $T_{c}$ in cuprate superconductors. The
on-site Coulomb repulsion $U$ takes value $2.44$, which is the
effective value near optimal doping in the hubbard-type
model~\cite{zhujx13,zhou32}. We study LA effect on
the patterns around impurities in DSCs near optimal doping with the
filling factor $n_{f}=\sum_{i\sigma}c_{i\sigma}^{
\dagger}c_{i\sigma}/(N_{x}N_{y})=0.85$ (i.e., the hole doping
$x=0.15$), where $N_{x}$, $N_{y}$ are the linear dimension of the
unit cell. The BdG equations are solved self-consistently for a
square lattice of $24\times24$ sites, and the periodic boundary
conditions are adopted. The numerical calculation is performed at a
very low temperature, $T=10^{-5}$K, to extract the low-energy
physics. The local effective field is taken to be $h_{eff}$ at the
impurity site and zero otherwise. The DSC order parameter at the
$i$th site is defined as
$\Delta_{i}=(\Delta_{i,i+e_{x}}+\Delta_{i,i-e_{x}}-\Delta_{i,i+e_{y}}-\Delta_{i,i-e_{y}})/4$,
and the spin order parameter is $M_{i}=(-1)^{i}m_{i}$.



\begin{figure}[!htb]
\center
\includegraphics[width=0.48\textwidth]{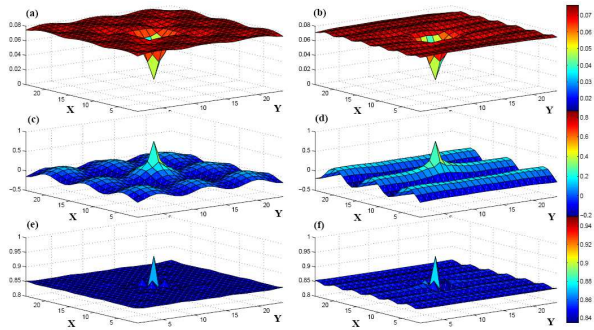}
\caption{(color online) The surface plots of orders around the
magnetic impurity on a unit cell of size $24\times24$ sites with and without lattice anisotropy. (a), (c), and (e) are the spatial
distributions of the DSC, spin and charge orders without LA. (b), (d), and (f)
are the same plots but the lattice
anisotropy ratio $\eta=t_{y}/t_{x}=0.99$. Here we take $U=2.44$, $x=0.15$,
$t^{\prime}=-0.25$, and $h_{eff}=3$.} \label{f.1}
\end{figure}

In FIG.~\ref{f.1}, we plot the spatial distributions of the DSC,
spin, and charge orders around the impurity site. As one can see 
[FIG.~\ref{f.1} (a), (c), and (e)],
all the three orders exhibit checkerboard modulations around the
magnetic impurity. Similar to the nonmagnetic impurity case, a SDW
with a period of 8$a$ checkerboard pattern is observed around the
magnetic impurity, which is in agreement with the NS
experiment~\cite{lake4}. However, one notes that the checkerboard
pattern of the DSC order can also be induced by the magnetic
impurity, and a weak associated CDW pattern is observed, which is
different from the nonmagnetic impurity case~\cite{zhujx13}.
Moreover, the modulated DSC and CDW orders share the same
periodicity 4$a$. On the whole, the DSC order has a local minimum at
the impurity site while the amplitudes of the CDW and SDW orders
reach global maxima. This is the common feature of orders around the
magnetic impurity despite various parameters. Therefore, in view of
these features, we clearly see the relationship of competition and
coexistence between antiferromagnetic and DSC orderings.


Below we discuss the effect of LA on the pattern around a single impurity
[See Fig.~\ref{f.1} (b), (d), and (f)].
As one can see, a symmetry-broken transition from
the checkerboard to stripe pattern is induced. In this case, even if
there is a very weak LA, this transition takes
place. The reason lies in the fact that LA
$t_{y}<t_{x}$ breaks the symmetry of the system, which leads to the
formation of stripes along y direction favorable. Therefore, we see
that LA is favorable to the formation of stripes in
DSCs. In addition, one notes that the anisotropy-induced SDW stripes
also show a modulation with periodicity 8$a$, which coexists with
the DSC order. This indicates that the 8$a$ modulations appearing in
the SDW order is a robust feature in spite of different parameters.
This conclusion is consistent with our previous study~\cite{zuo28}.

\begin{figure}[!htb]
\center
\includegraphics[width=0.48\textwidth]{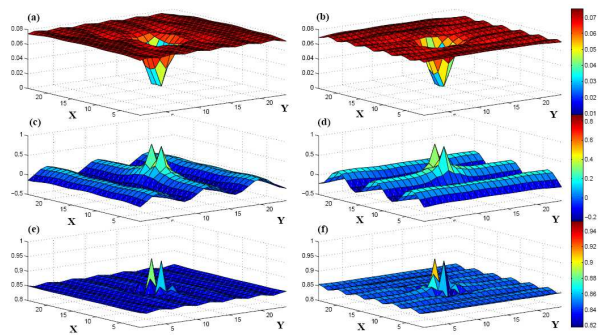}
\caption{(color online) The surface plots of orders around two
magnetic impurities in the presence of LA. (a), (c), and (e) are plots
in the case with $\eta=0.983$, while (b), (d), and (f) with $\eta=0.981$.} \label{f.2}
\end{figure}

Now let us turn to consider the two-impurity case. Without LA, an x-orientated stripe structure can be induced due to
the pinning effect of the magnetic impurities which have been placed
along x direction [Not shown but similar to FIG.~\ref{f.2} (a), (c), and (e)].
Similar to the single
impurity case, the impurity-pinned SDW stripe also show a modulation
with periodicity 8$a$, while the coexisting DSC and CDW stripes
share the same periodicity 4$a$. For the case of two magnetic
impurities with LA present, one could expect that
the impurity-pinned effect would compete with LA
effect, leading to orientation transition from the longitudinal (x
direction) impurity-pinned stripe into the transverse (y direction)
pattern when the LA ratio reaches some critical
value. This speculation is confirmed by the following calculations,
and the critical value is found to be $\eta^{*}\sim0.982$, which is
close to the above estimated value
$t_{y}/t_{x}\sim(b/a)^{-4}\sim0.9615$.

Accordingly, the ratio of
lattice constants is estimated as $b/a\sim1.005$. Thus, we see that
only a weak LA is needed to observe this pattern
transition in the disorder-pinned stripe 


phase in DSCs experimentally. If adopting
the estimated value by Normand and Kampf or by Citro and
Marinaro~\cite{normand19,citro33}, one gets
$\eta=t_{y}/t_{x}=|cos(\pi-2\Phi)|\sim0.9848$, with an angular
distortion angle $\Phi=5^{\circ}$. As shown in FIG.~\ref{f.2} (a), (c) and (e), when
the LA ratio is larger than the critical value (a
weak anisotropy), $\eta=0.983>\eta^{*}$, the x-orientated stripe
structure is still stable. This case is very similar
to the case without LA, except that the modulation
amplitudes are slightly larger than those in the latter case.
However, for the relatively strong anisotropy with
$\eta=0.981<\eta^{*}$, the orientation transition from x-stripe to
y-stripe pattern takes place, as shown in FIG.~\ref{f.2} (b), (d), and (f). Clearly,
in this case, LA effect dominates over the impurity
pinning effect, and the y-stripe pattern is energetically more
favorable. Also, as shown by all those stripe plots, the
checkerboard-like modulations are still visible, especially in the
DSC and SDW orders, which originate from the effect of QPI. Thus, we
see that one can produce different modulated patterns by inserting
magnetic impurities or tuning the LA in DSCs while
the modulation periodicity holds the same.


For comparison, we also consider the cases with various strength of
effective field, $h_{eff}=1$ or $10$, and with a long distance
between the two impurities. In these cases, we obtain similar
results, and reach the same conclusion as discussed above, except
that the critical value $\eta^{*}$ changes more or less. For the
cases with $h_{eff}=1$ and with a 8$a$ distance between the two
impurities, we happen to get the exact critical values of the
LA ratio, i.e., $\eta^{*}=0.984$ and $0.998$, which are
both larger than the previous case with $h_{eff}=3$. Meanwhile, for
the case with $h_{eff}=10$, the corresponding critical values is
found to be $\eta^{*}\sim0.981$, which is smaller than the above
cases. The reason lies in the fact that for a weaker effective field
or a longer distance, the impurity pinning effect is reduced, while
an opposite tendency is produced by a stronger effective field.


\begin{figure}[!htb]
\center
\includegraphics[width=0.48\textwidth]{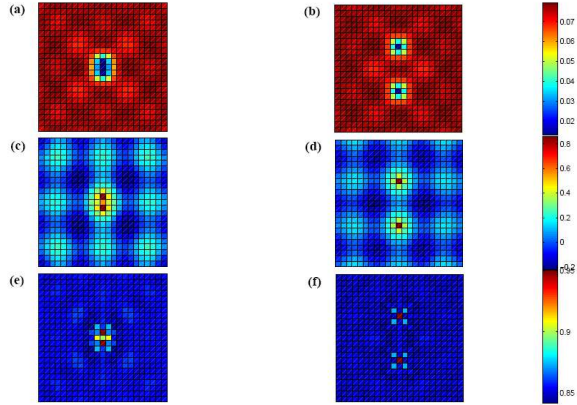}
\caption{(color online) The surface plots of orders around two
magnetic impurities at the critical
LA ratio with different distances between the two impurities.
(a), (c), and (e) are plots
in the case with $\eta^{*}=0.984$, while (b), (d), and (f) the case with $\eta^{*}=0.998$.} \label{f.3}
\end{figure}

At the exact critical value of the
LA, it is found that the structures around magnetic
impurities restore checkerboard patterns because the competition
between the impurity pinning effect and LA reaches
the balance point, as shown in FIG.~\ref{f.3}. In
this case, it seems that the two impurities could induce their own
checkerboard patterns independently, without interference with each
other. These results indicate that one could observe the
stripe-checkerboard-stripe transition under some conditions by
tuning $\eta$ in impurity-substituted DSCs.



\quad In summery, we have studied the effect of LA on the
patterns around magnetic impurities based on the
$t_{x}-t_{y}-t^{\prime}-U-V$ model. It is demonstrated that LA could 
induce pattern transition around impurities in
near-optimally doped DSCs. In the single impurity case, it is found
that even a very weak LA could induce a transition
from the checkerboard to stripe pattern, because the symmetry of the
system is broken. Modulated SDW pattern with 8$a$ periodicity is
observed, which coincides with the NS data. Meanwhile, the modulated
DSC and CDW orders share the same periodicity 4$a$. For the
two-impurity case, a transition from the x-direction impurity-pinned
stripe into the y-direction stripe pattern is observed as the
LA ratio reaches the critical value $\eta^{*}$. At
the exact critical values, it is found that the structures around
magnetic impurities could restore checkerboard patterns. It is
expected that these phenomena could be observed in the STM and NS
experiments on DSCs under varying pressure or temperature.

This work is supported by the National Nature
Science Foundation of China (Grants No. 10904063 and No. 10804047). CDG would also like to thank the 973 Project
(Project No. 2006CB601002).

\end{document}